\documentclass[12pt,a4paper]{article}
\usepackage{amsmath}
\usepackage{latexsym}
\makeatletter
\def\rddots{\mathinner{\mkern1mu\raise\p@%
    \vbox{\kern7\p@\hbox{.}}\mkern2mu%
    \raise4\p@\hbox{.}\mkern2mu\raise7\p@\hbox{.}\mkern1mu}}
\makeatother
\newcommand{\fukuso}{{\mathbf C}}
\newcommand{\futon}{{\bf N}}

\begin{document}

\title{\sl How to Calculate the Exponential of Matrices}
\author{
  Kazuyuki FUJII
  \thanks{E-mail address : fujii@yokohama-cu.ac.jp }\quad and\ 
  Hiroshi OIKE
  \thanks{E-mail address : oike@tea.ocn.ne.jp }\\
  ${}^{*}$Department of Mathematical Sciences\\
  Yokohama City University\\
  Yokohama, 236--0027\\
  Japan\\
  ${}^{\dagger}$Takado\ 85--5,\ Yamagata, 990--2464\\
  Japan\\
  }
\date{}
\maketitle
\begin{abstract}
  How to calculate the exponential of matrices in an explicit manner is 
  one of fundamental problems in almost all subjects in Science.
  
  Especially in Mathematical Physics or Quantum Optics many problems are 
  reduced to this calculation by making use of some approximations whether 
  they are appropriate or not. However, it is in general not easy.
  
  In this paper we give a very useful formula which is both elementary and 
  getting on with computer.
\end{abstract}
%

%\newpage

%
%
%     Honbun
%
%
\newpage
To calculate the exponential of matrices 
\begin{equation}
\label{eq:the exponential}
\mbox{e}^{A}=\sum_{m=0}^{\infty}\frac{A^{m}}{m!}
\quad \mbox{for}\ A\ \in\ M(n;\fukuso)
\end{equation}
is one of fundamental problems in almost all subjects in Science \footnote{ 
See the dictionary \cite{MSJ} concerning several mathematical notations 
in the paper}. 

In fact, in Mathematical Physics or Quantum Optics many problems are reduced 
to this calculation by use of some approximations. 
However, this is a very hard problem. See, for example, hard and ``painful" 
calculations in \cite{FHKW} and \cite{KF4}.

In usual textbooks of Linear Algebra (see for example \cite{IS}) 
it is recommended to diagonalize $A$ like 
\begin{equation}
\label{eq:diagonalization}
A=UD_{A}U^{-1}\qquad \mbox{for\ some\ }\ U\ \in\ GL(n;\fukuso)
\end{equation}
where $D_{A}$ is the diagonal matrix consisting of all eigenvalues of 
$A$. Unfortunately, this method is not realistic as known well.

Let us explain in more detail. First of all we write the characteristic 
equation (polynomial) of $A$ :
\begin{equation}
\label{eq:characteristic equation}
0=|\lambda E-A|
 =\lambda^{n}+p_{1}\lambda^{n-1}+\cdots +p_{n-1}\lambda+p_{n}
\end{equation}
where $p_{1}=-\mbox{tr}A$ and $p_{n}=(-1)^{n}\mbox{det}(A)$. 
(\ref{eq:characteristic equation}) can be decomposed into
\begin{equation}
\label{eq:decomposition}
\lambda^{n}+p_{1}\lambda^{n-1}+\cdots +p_{n-1}\lambda+p_{n}
=(\lambda -\alpha_{1})(\lambda -\alpha_{2})\cdots (\lambda -\alpha_{n})
\end{equation}
where $\alpha_{j} \in \fukuso$. From this we have
\begin{eqnarray}
\label{eq:relations}
p_{1}&=&-\sum_{j=1}^{n}\alpha_{j} \nonumber \\
p_{2}&=&(-1)^{2}\sum_{i<j}^{n}\alpha_{i}\alpha_{j} \nonumber \\
&\vdots& \nonumber \\
p_{n}&=&(-1)^{n}\prod_{j=1}^{n}\alpha_{j}
\end{eqnarray}

For the eigenvalue $\alpha_{j}\ (j=1,2,\cdots,n)$ we define $|\alpha_{j}) \in 
\fukuso^{n}$ to be the eigenvector which is not necessarily normalized 
\footnote{In general, to normalize a system of vectors is not easy. 
See the example in the latter half.}
\begin{equation}
\label{eq:eigen-vector}
A|\alpha_{j})=\alpha_{j}|\alpha_{j}),
\end{equation}
then we obtain the diagonalization (\ref{eq:diagonalization}) if we define 
$U$ as
\begin{equation}
\label{eq:general-linear}
U=\left(|\alpha_{1}),|\alpha_{2}),\cdots,|\alpha_{n})\right)\ 
\in\ GL(n;\fukuso)
\end{equation}
and $D_{A}=\mbox{diag}(\alpha_{1},\alpha_{2},\cdots,\alpha_{n})$. 

Weak points of this method are that
\par \noindent
(a)\ to find the eigenvectors (\ref{eq:eigen-vector}) explicitly, 
\par \noindent
(b)\ to calculate $U^{-1}$ the inverse of $U$.
\par \noindent
They become more and more difficult as $n$ becomes large.

\par \vspace{5mm} 
Now, it is better to change the strategy. The famous theorem of 
Cayley--Hamilton gives
\begin{equation}
\label{eq:cayley-hamilton}
A^{n}+p_{1}A^{n-1}+\cdots +p_{n-1}A+p_{n}E=0\ \Longrightarrow 
A^{n}=-p_{1}A^{n-1}-\cdots -p_{n-1}A-p_{n}E
\end{equation}
, so we can formally write 
\begin{equation}
\label{eq:expansion}
\mbox{e}^{A}=\sum_{m=0}^{\infty}\frac{A^{m}}{m!}
=f_{0}E+f_{1}A+f_{2}A^{2}+\cdots +f_{n-1}A^{n-1}
\end{equation}
where $f_{j}=f_{j}(p_{1},p_{2},\cdots, p_{n})$ for simplicity. 
{\bf The purpose in the following is to determine 
$\{f_{0},f_{1},\cdots,f_{n-1}\}$ explicitly} \footnote{This way of thinking 
is very natural because $A$ is finite--dimensional}.

Here we use the notation like
\begin{eqnarray}
\label{eq:convenient notation}
&&f_{0}E+f_{1}A+f_{2}A^{2}+\cdots +f_{n-1}A^{n-1} \nonumber \\
&\ =&Ef_{0}+Af_{1}+A^{2}f_{2}+\cdots +A^{n-1}f_{n-1}
\equiv \left(E,A,A^{2},\cdots,A^{n-1}\right)
\left(
\begin{array}{ccccc}
 f_{0}\\
 f_{1}\\
 f_{2}\\
 \vdots\\
 f_{n-1}
\end{array}
\right).
\end{eqnarray}
Note that 
\[
A^{n}=\left(E,A,A^{2},\cdots,A^{n-1}\right)
\left(
\begin{array}{ccccc}
 -p_{n}\\
 -p_{n-1}\\
 \vdots\\
 -p_{2}\\
 -p_{1}
\end{array}
\right)
\]
from (\ref{eq:cayley-hamilton}).

From (\ref{eq:characteristic equation}) we define the matrix
\begin{equation}
L=
\left(
  \begin{array}{cccccc}
  0 & 0 & 0 & \cdots & 0 & -p_{n}   \\
  1 & 0 & 0 & \cdots & 0 & -p_{n-1} \\ 
  0 & 1 & 0 & \cdots & 0 & -p_{n-2} \\ 
  \vdots & \vdots  & \vdots &  & \vdots & \vdots \\
  0 & 0 & 0 & \cdots & 0 & -p_{2}   \\
  0 & 0 & 0 & \cdots & 1 & -p_{1}   
  \end{array}
\right),
\end{equation}
which is called the companion matrix. 
It is well known that $L$ also satisfies the same characteristic 
equation (\ref{eq:characteristic equation})
\begin{equation}
\label{eq:characteristic equation 2}
0=|\lambda E-L|
 =\lambda^{n}+p_{1}\lambda^{n-1}+\cdots +p_{n-1}\lambda+p_{n}.
\end{equation}
We reconsider an important role that $L$ plays.

Now, we are in a position to state the main result.
\par \vspace{3mm} \noindent
{\bf Fundamental Lemma}\quad For any $m \in \futon\cup \{0\}$ we have
\begin{equation}
\label{eq:main lemma}
A^{m}=\left(E,A,A^{2},\cdots,A^{n-1}\right)L^{m}{\mathbf e}_{1}
\end{equation}
where ${\mathbf e}_{1}=(1,0,0,\cdots,0)^{\mbox{T}}$ (see the notation in 
(\ref{eq:convenient notation}) once more). 

\par \noindent
The proof is by mathematical induction. 

\par \vspace{5mm}
This leads us to
\begin{equation}
\label{eq:main equation}
\mbox{e}^{A}=\left(E,A,A^{2},\cdots,A^{n-1}\right)
\mbox{e}^{L}{\mathbf e}_{1}.
\end{equation}
That is, the calculation of $\mbox{e}^{A}$ is reduced to that of 
$\mbox{e}^{L}$. We must again calculate the exponential ! 
What is interesting ? In this case, we can make $L$ diagonal completely 
because it is simple enough \footnote{There is another diagonalization 
(communicated by T. Suzuki)}.

Let us solve the equation(s)
\begin{equation}
\label{eq:eigen equation}
L|\alpha_{j})=\alpha_{j}|\alpha_{j})\quad \mbox{for}\quad j=1\sim n\ , 
\end{equation}
which is easily obtained to be 
\begin{equation}
|\alpha)=
\left(
\begin{array}{ccccc}
 p_{n-1}+p_{n-2}\alpha+\cdots+p_{1}\alpha^{n-2}+\alpha^{n-1}\\
 p_{n-2}+p_{n-3}\alpha+\cdots+p_{1}\alpha^{n-3}+\alpha^{n-2}\\
 \vdots\\
 p_{1}+\alpha\\
 1
\end{array}
\right),
\end{equation}
where $\alpha=\alpha_{1},\alpha_{2},\cdots,\alpha_{n}$ for simplicity. 
It is of course $(\alpha_{i}|\alpha_{j})\ne \delta_{ij}$. 
Then we have
\begin{equation}
\label{eq:general-linear 2}
U_{L}=\left(|\alpha_{1}),|\alpha_{2}),\cdots,|\alpha_{n})\right)\ \in\ 
GL(n;\fukuso)
\end{equation}
and
\begin{equation}
\label{eq:diagonalization 2}
L=U_{L}D_{A}{U_{L}}^{-1}\ \Longrightarrow\ 
\mbox{e}^{L}=U_{L}\mbox{e}^{D_{A}}{U_{L}}^{-1}.
\end{equation}

At first sight, it seems difficult to calculate ${U_{L}}^{-1}$. However, 
we have a simple decomposition of $U_{L}$ like 
\begin{equation}
\label{eq:decomposition of U}
U_{L}=P_{L}Q_{L}
\end{equation}
where $P_{L}$ and $Q_{L}$ are given respectively as

\begin{eqnarray}
\label{eq:P}
&&P_{L}=
\left(
  \begin{array}{cccccc}
    p_{n-1} & p_{n-2} & p_{n-3} & \cdots & p_{1} & 1         \\
    p_{n-2} & p_{n-3} & \cdots & p_{1} & 1 & 0               \\
    p_{n-3} & \vdots  & \rddots & \rddots & \vdots & \vdots  \\
    \vdots & p_{1} & \rddots & 0 & 0 & 0                     \\
    p_{1} & 1 & \cdots & 0 & 0 & 0                           \\
    1 & 0 & \cdots & 0 & 0 & 0 
  \end{array}            
\right),  \\
\label{eq:Q}
&&Q_{L}=
\left(
  \begin{array}{cccccc}
  1 & 1 & \cdots &  & 1 & 1  \\
  \alpha_{1} & \alpha_{2} & \cdots & & \alpha_{n-1} & \alpha_{n} \\
  \alpha_{1}^{2} & \alpha_{2}^{2} & \cdots & & \alpha_{n-1}^{2} & 
  \alpha_{n}^{2} \\
  \vdots & \vdots &  &  & \vdots & \vdots \\
  \alpha_{1}^{n-2} & \alpha_{2}^{n-2} & \cdots & & \alpha_{n-1}^{n-2} & 
  \alpha_{n}^{n-2} \\
  \alpha_{1}^{n-1} & \alpha_{2}^{n-1} & \cdots & & \alpha_{n-1}^{n-1} & 
  \alpha_{n}^{n-1} 
  \end{array}
\right).
\end{eqnarray}
Therefore $U_{L}^{-1}=Q_{L}^{-1}P_{L}^{-1}$. Must we calculate 
both $Q_{L}^{-1}$ and $P_{L}^{-1}$ again ?  From (\ref{eq:main equation}) 
it is just $U_{L}^{-1}{\mathbf e}_{1}$ not $U_{L}^{-1}$ itself that we must 
calculate. It is easy to see
\begin{equation}
U_{L}^{-1}{\mathbf e}_{1}=Q_{L}^{-1}P_{L}^{-1}{\mathbf e}_{1}
=Q_{L}^{-1}{\mathbf e}_{n}
\end{equation}
where ${\mathbf e}_{n}=(0,0,\cdots,0,1)^{\mbox{T}}$ and 
$Q_{L}^{-1}{\mathbf e}_{n}$ is given as
\begin{equation}
Q_{L}^{-1}{\mathbf e}_{n}
=
\frac{1}{|Q_{L}|}
\left(
  \begin{array}{c}
    \mbox{the cofactor of}\ \alpha_{1}^{n-1} \\
    \mbox{the cofactor of}\ \alpha_{2}^{n-1} \\
    \vdots \\
    \mbox{the cofactor of}\ \alpha_{n-1}^{n-1} \\
    \mbox{the cofactor of}\ \alpha_{n}^{n-1} \\
  \end{array}            
\right)
=(-1)^{n+1}
\left(
  \begin{array}{c}
    \frac{1}{\prod_{j=1,j\ne 1}^{n}(\alpha_{j}-\alpha_{1})} \\
    \frac{1}{\prod_{j=1,j\ne 2}^{n}(\alpha_{j}-\alpha_{2})} \\
    \vdots \\
    \frac{1}{\prod_{j=1,j\ne n-1}^{n}(\alpha_{j}-\alpha_{n-1})} \\
    \frac{1}{\prod_{j=1,j\ne n}^{n}(\alpha_{j}-\alpha_{n})} \\
  \end{array}            
\right). 
\end{equation}
Moreover, we can determine $U_{L}$ completely. From (\ref{eq:relations}) we 
define
\begin{equation}
\left(p_{j}\right)_{k}=p_{j} - \left\{\mbox{all the terms containing}\ 
\alpha_{k}\right\}\quad \mbox{for}\quad 1\leq k\leq n.
\end{equation}
For example, 
\begin{eqnarray*}
&&\left(p_{1}\right)_{1}=-(\alpha_{2}+\cdots + \alpha_{n}),\quad 
\left(p_{n-1}\right)_{1}=(-1)^{n-1}\alpha_{2}\cdots \alpha_{n}, \\
&&\left(p_{1}\right)_{2}=-(\alpha_{1}+\alpha_{3}+\cdots + \alpha_{n}),\quad 
\left(p_{n-1}\right)_{2}=(-1)^{n-1}\alpha_{1}\alpha_{3}\cdots \alpha_{n}.
\end{eqnarray*}
Then 
\begin{equation}
\label{eq:}
U_{L}
=\left(|\alpha_{1}),|\alpha_{2}),\cdots,|\alpha_{n})\right)
=
\left(
  \begin{array}{ccccc}
  \left(p_{n-1}\right)_{1} & \left(p_{n-1}\right)_{2} & \cdots & 
  \left(p_{n-1}\right)_{n-1} & \left(p_{n-1}\right)_{n}           \\
  \left(p_{n-2}\right)_{1} & \left(p_{n-2}\right)_{2} & \cdots & 
  \left(p_{n-2}\right)_{n-1} & \left(p_{n-2}\right)_{n}           \\
  \vdots & \vdots & \ddots & \vdots & \vdots                      \\
  \left(p_{1}\right)_{1} & \left(p_{1}\right)_{2} & \cdots & 
  \left(p_{1}\right)_{n-1} & \left(p_{1}\right)_{n}               \\
  1 & 1 & \cdots & 1 & 1
  \end{array}            
\right).
\end{equation}
Noting
\begin{eqnarray}
&&\mbox{e}^{L}{\mathbf e}_{1}
=U_{L}\mbox{e}^{D_{A}}{U_{L}}^{-1}{\mathbf e}_{1} \nonumber \\
&=&(-1)^{n+1}
\left(
  \begin{array}{ccccc}
  \left(p_{n-1}\right)_{1} & \left(p_{n-1}\right)_{2} & \cdots & 
  \left(p_{n-1}\right)_{n-1} & \left(p_{n-1}\right)_{n}           \\
  \left(p_{n-2}\right)_{1} & \left(p_{n-2}\right)_{2} & \cdots & 
  \left(p_{n-2}\right)_{n-1} & \left(p_{n-2}\right)_{n}           \\
  \vdots & \vdots & \ddots & \vdots & \vdots                      \\
  \left(p_{1}\right)_{1} & \left(p_{1}\right)_{2} & \cdots & 
  \left(p_{1}\right)_{n-1} & \left(p_{1}\right)_{n}               \\
  1 & 1 & \cdots & 1 & 1
  \end{array}            
\right)
\left(
\begin{array}{c}
\frac{\mbox{e}^{\alpha_{1}}}{\prod_{j=1,j\ne 1}^{n}(\alpha_{j}-\alpha_{1})} \\
\frac{\mbox{e}^{\alpha_{2}}}{\prod_{j=1,j\ne 2}^{n}(\alpha_{j}-\alpha_{2})} \\
\vdots \\
\frac{\mbox{e}^{\alpha_{n-1}}}{\prod_{j=1,j\ne n-1}^{n}
(\alpha_{j}-\alpha_{n-1})} \\
\frac{\mbox{e}^{\alpha_{n}}}{\prod_{j=1,j\ne n}^{n}(\alpha_{j}-\alpha_{n})} \\
\end{array}            
\right)  \nonumber \\
&&{}
\end{eqnarray}
we finally obtain 
\[
\mbox{e}^{A}=\left(E,A,A^{2},\cdots,A^{n-1}\right)
\mbox{e}^{L}{\mathbf e}_{1}
\equiv f_{0}E+f_{1}A+f_{2}A^{2}+\cdots +f_{n-1}A^{n-1}
\]
with
\begin{equation}
f_{l}=(-1)^{n+1}\sum_{k=1}^{n}
\frac{\left(p_{n-l-1}\right)_{k}\mbox{e}^{\alpha_{k}}}{\prod_{j=1,j\ne k}^{n}
(\alpha_{j}-\alpha_{k})}\qquad 0\leq l\leq n-1
\end{equation}
in a complete manner \footnote{The result is also obtained by the spectral 
decomposition of matrices (communicated by T. Suzuki). However, it seems to us 
that the method is much elementary. The result has not been written in 
standard textbooks in Linear Algebra as far as we know}.

\par \noindent
That is, our calculation is based on only simple operations like the powers of 
matrices or sum of them, etc, which is easily performed by computer.

\par \vspace{5mm}
Let us list some important examples (the case of $n=3$ and $4$).

\par \vspace{3mm} \noindent
\underline{{\bf $n=3$}}\quad For
\begin{equation}
A=
\left(
  \begin{array}{ccc}
    a_{11} & a_{12} & a_{13} \\
    a_{21} & a_{22} & a_{23} \\
    a_{31} & a_{32} & a_{33} 
  \end{array}            
\right)\ \in\ M(3;\fukuso)
\end{equation}
we have
\begin{eqnarray}
0&=&|\lambda E-A|  \nonumber \\
 &=&\lambda^{3}-(a_{11}+a_{22}+a_{33})\lambda^{2}+
    (a_{11}a_{22}+a_{11}a_{33}+a_{22}a_{33}-
     a_{12}a_{21}-a_{13}a_{31}-a_{23}a_{32})\lambda -  \nonumber \\
 &{}&
 (a_{11}a_{22}a_{33}+a_{12}a_{23}a_{31}+a_{13}a_{21}a_{32}
 -a_{13}a_{22}a_{31}-a_{12}a_{21}a_{33}-a_{11}a_{23}a_{32}).
\end{eqnarray}
The Cardano formula (see for example \cite{KF3}) gives three solutions 
$\{\alpha_{1},\alpha_{2},\alpha_{3}\}$. 
Then 
\begin{equation}
p_{1}=-(\alpha_{1}+\alpha_{2}+\alpha_{3}),\quad
p_{2}=\alpha_{1}\alpha_{2}+\alpha_{1}\alpha_{3}+\alpha_{2}\alpha_{3},\quad
p_{3}=-\alpha_{1}\alpha_{2}\alpha_{3}
\end{equation}
and
\begin{equation}
L=
\left(
  \begin{array}{ccc}
    0 & 0 & -p_{3} \\
    1 & 0 & -p_{2} \\
    0 & 1 & -p_{1}
  \end{array}            
\right)=U_{L}D_{A}{U_{L}}^{-1}
\end{equation}
and
\begin{eqnarray}
U_{L}
&=&
\left(
  \begin{array}{ccc}
    p_{2}+p_{1}\alpha_{1}+\alpha_{1}^{2} & 
    p_{2}+p_{1}\alpha_{2}+\alpha_{2}^{2} & 
    p_{2}+p_{1}\alpha_{3}+\alpha_{3}^{2}   \\
    p_{1}+\alpha_{1} & p_{1}+\alpha_{2} & p_{1}+\alpha_{3} \\
    1 & 1 & 1
  \end{array}            
\right) \nonumber \\
&=&
\left(
  \begin{array}{ccc}
   \alpha_{2}\alpha_{3} & \alpha_{1}\alpha_{3} & \alpha_{1}\alpha_{2} \\
  -(\alpha_{2}+\alpha_{3}) & -(\alpha_{1}+\alpha_{3}) & 
  -(\alpha_{1}+\alpha_{2}) \\      
  1 & 1 & 1
  \end{array}
\right)
\end{eqnarray}
and
\begin{equation}
\left(
\begin{array}{c}
 f_{0} \\
 f_{1} \\
 f_{2}
\end{array}
\right)
=\mbox{e}^{L}{\mathbf e}_{1}
=
\left(
  \begin{array}{ccc}
   \alpha_{2}\alpha_{3} & \alpha_{1}\alpha_{3} & \alpha_{1}\alpha_{2} \\
  -(\alpha_{2}+\alpha_{3}) & -(\alpha_{1}+\alpha_{3}) & 
  -(\alpha_{1}+\alpha_{2}) \\
   1 & 1 & 1
  \end{array}
\right)
\left(
\begin{array}{c}
\frac{\mbox{e}^{\alpha_{1}}}{(\alpha_{2}-\alpha_{1})(\alpha_{3}-\alpha_{1})}\\
\frac{\mbox{e}^{\alpha_{2}}}{(\alpha_{1}-\alpha_{2})(\alpha_{3}-\alpha_{2})}\\
\frac{\mbox{e}^{\alpha_{3}}}{(\alpha_{1}-\alpha_{3})(\alpha_{2}-\alpha_{3})}
\end{array}
\right).
\end{equation}

Therefore we have finally
\begin{equation}
\mbox{e}^{A}=f_{0}E+f_{1}A+f_{2}A^{2}
\end{equation}
with
\begin{eqnarray}
f_{0}&=&
\frac{\alpha_{2}\alpha_{3}\mbox{e}^{\alpha_{1}}}
{(\alpha_{2}-\alpha_{1})(\alpha_{3}-\alpha_{1})}
+
\frac{\alpha_{1}\alpha_{3}\mbox{e}^{\alpha_{2}}}
{(\alpha_{1}-\alpha_{2})(\alpha_{3}-\alpha_{2})}
+
\frac{\alpha_{1}\alpha_{2}\mbox{e}^{\alpha_{3}}}
{(\alpha_{1}-\alpha_{3})(\alpha_{2}-\alpha_{3})}, \\
f_{1}&=&
-\frac{(\alpha_{2}+\alpha_{3})\mbox{e}^{\alpha_{1}}}
{(\alpha_{2}-\alpha_{1})(\alpha_{3}-\alpha_{1})}
-
\frac{(\alpha_{1}+\alpha_{3})\mbox{e}^{\alpha_{2}}}
{(\alpha_{1}-\alpha_{2})(\alpha_{3}-\alpha_{2})}
-
\frac{(\alpha_{1}+\alpha_{2})\mbox{e}^{\alpha_{3}}}
{(\alpha_{1}-\alpha_{3})(\alpha_{2}-\alpha_{3})}, \\
f_{2}&=&
\frac{\mbox{e}^{\alpha_{1}}}
{(\alpha_{2}-\alpha_{1})(\alpha_{3}-\alpha_{1})}
+
\frac{\mbox{e}^{\alpha_{2}}}
{(\alpha_{1}-\alpha_{2})(\alpha_{3}-\alpha_{2})}
+
\frac{\mbox{e}^{\alpha_{3}}}
{(\alpha_{1}-\alpha_{3})(\alpha_{2}-\alpha_{3})}.
\end{eqnarray}

\par \vspace{10mm} \noindent
\underline{{\bf $n=4$}}\quad For
\begin{equation}
A=
\left(
  \begin{array}{cccc}
    a_{11} & a_{12} & a_{13} & a_{14} \\
    a_{21} & a_{22} & a_{23} & a_{24} \\
    a_{31} & a_{32} & a_{33} & a_{34} \\
    a_{41} & a_{42} & a_{43} & a_{44} 
  \end{array}            
\right)\ \in\ M(4;\fukuso)
\end{equation}
we have
\begin{eqnarray}
0&=&|\lambda E-A|  \nonumber \\
 &=&\lambda^{4}-(a_{11}+a_{22}+a_{33}+a_{44})\lambda^{3}+
    (a_{11}a_{22}+a_{11}a_{33}+a_{11}a_{44}+a_{22}a_{33}
\nonumber \\
  &&+a_{22}a_{44}+a_{33}a_{44}-a_{12}a_{21}-a_{13}a_{31}
    -a_{14}a_{41}-a_{23}a_{32}-a_{24}a_{42}-a_{34}a_{43})\lambda^{2}  
\nonumber \\
&{}&
-(a_{11}a_{22}a_{33}+a_{11}a_{22}a_{44}
 +a_{11}a_{33}a_{44}+a_{22}a_{33}a_{44}
 +a_{12}a_{23}a_{31}+a_{12}a_{24}a_{41}
\nonumber \\
&{}&
 +a_{13}a_{21}a_{32}+a_{13}a_{34}a_{41}
 +a_{14}a_{21}a_{42}+a_{14}a_{31}a_{43}
 +a_{23}a_{34}a_{42}+a_{24}a_{32}a_{43}
\nonumber \\
&{}&
 -a_{11}a_{23}a_{32}-a_{11}a_{24}a_{42}
 -a_{11}a_{34}a_{43}-a_{12}a_{21}a_{33}
 -a_{12}a_{21}a_{44}-a_{13}a_{22}a_{31}
\nonumber \\
&{}&
 -a_{13}a_{31}a_{44}-a_{14}a_{22}a_{41}
 -a_{14}a_{33}a_{41}-a_{22}a_{34}a_{43}-a_{23}a_{32}a_{44}-a_{24}a_{33}a_{42}
 )\lambda   \nonumber \\
&{}&
+\det(A)
\end{eqnarray}
where $\det(A)$ is omitted, see for example \cite{IS}.\quad 
The Ferrari formula or Euler one (see \cite{KF3}) gives four solutions 
$\{\alpha_{1},\alpha_{2},\alpha_{3},\alpha_{4}\}$ \footnote{They are of course 
too complicated}. 
Then 
\begin{eqnarray}
&&p_{1}=-(\alpha_{1}+\alpha_{2}+\alpha_{3}+\alpha_{4}),\quad
p_{2}=\alpha_{1}\alpha_{2}+\alpha_{1}\alpha_{3}+\alpha_{1}\alpha_{4}+
\alpha_{2}\alpha_{3}+\alpha_{2}\alpha_{4}+\alpha_{3}\alpha_{4},
\nonumber \\
&&p_{3}=-(\alpha_{1}\alpha_{2}\alpha_{3}+\alpha_{1}\alpha_{2}\alpha_{4}+
\alpha_{1}\alpha_{3}\alpha_{4}+\alpha_{2}\alpha_{3}\alpha_{4}),\quad 
p_{4}=\alpha_{1}\alpha_{2}\alpha_{3}\alpha_{4}
\end{eqnarray}
and
\begin{small}
\begin{equation}
L=
\left(
  \begin{array}{cccc}
    0 & 0 & 0 & -p_{4} \\
    1 & 0 & 0 & -p_{3} \\
    0 & 1 & 0 & -p_{2} \\
    0 & 0 & 1 & -p_{1}
  \end{array}            
\right)=U_{L}D_{A}{U_{L}}^{-1}
\end{equation}
and
\begin{eqnarray}
&&\quad U_{L}= \nonumber \\
&&\left(
  \begin{array}{cccc}
    p_{3}+p_{2}\alpha_{1}+p_{1}\alpha_{1}^{2}+\alpha_{1}^{3} &
    p_{3}+p_{2}\alpha_{2}+p_{1}\alpha_{2}^{2}+\alpha_{2}^{3} &
    p_{3}+p_{2}\alpha_{3}+p_{1}\alpha_{3}^{2}+\alpha_{3}^{3} &
    p_{3}+p_{2}\alpha_{4}+p_{1}\alpha_{4}^{2}+\alpha_{4}^{3} \\
    p_{2}+p_{1}\alpha_{1}+\alpha_{1}^{2}&
    p_{2}+p_{1}\alpha_{2}+\alpha_{2}^{2}&
    p_{2}+p_{1}\alpha_{3}+\alpha_{3}^{2}&
    p_{2}+p_{1}\alpha_{4}+\alpha_{4}^{2} \\
    p_{1}+\alpha_{1} & p_{1}+\alpha_{2} & 
    p_{1}+\alpha_{3} & p_{1}+\alpha_{4} \\
    1 & 1 & 1 & 1
  \end{array}            
\right) \nonumber \\
&&\ =
\left(
  \begin{array}{cccc}
  -\alpha_{2}\alpha_{3}\alpha_{4} &
  -\alpha_{1}\alpha_{3}\alpha_{4} &
  -\alpha_{1}\alpha_{2}\alpha_{4} &
  -\alpha_{1}\alpha_{2}\alpha_{3} \\
   \alpha_{2}\alpha_{3}+\alpha_{2}\alpha_{4}+\alpha_{3}\alpha_{4} &
   \alpha_{1}\alpha_{3}+\alpha_{1}\alpha_{4}+\alpha_{3}\alpha_{4} &
   \alpha_{1}\alpha_{2}+\alpha_{1}\alpha_{4}+\alpha_{2}\alpha_{4} &
   \alpha_{1}\alpha_{2}+\alpha_{1}\alpha_{3}+\alpha_{2}\alpha_{3} \\
  -(\alpha_{2}+\alpha_{3}+\alpha_{4}) & 
  -(\alpha_{1}+\alpha_{3}+\alpha_{4}) & 
  -(\alpha_{1}+\alpha_{2}+\alpha_{4}) & 
  -(\alpha_{1}+\alpha_{2}+\alpha_{3}) \\      
  1 & 1 & 1 & 1
  \end{array}
\right) \nonumber \\
{}&&
\end{eqnarray}
\end{small}
and
\begin{equation}
\left(
\begin{array}{c}
 f_{0} \\
 f_{1} \\
 f_{2} \\
 f_{3}
\end{array}
\right)
=\mbox{e}^{L}{\mathbf e}_{1}
=
-U_{L}
\left(
\begin{array}{c}
\frac{\mbox{e}^{\alpha_{1}}}
{(\alpha_{2}-\alpha_{1})(\alpha_{3}-\alpha_{1})(\alpha_{4}-\alpha_{1})} \\
\frac{\mbox{e}^{\alpha_{2}}}
{(\alpha_{1}-\alpha_{2})(\alpha_{3}-\alpha_{2})(\alpha_{4}-\alpha_{2})} \\
\frac{\mbox{e}^{\alpha_{3}}}
{(\alpha_{1}-\alpha_{3})(\alpha_{2}-\alpha_{3})(\alpha_{4}-\alpha_{3})} \\
\frac{\mbox{e}^{\alpha_{4}}}
{(\alpha_{1}-\alpha_{4})(\alpha_{2}-\alpha_{4})(\alpha_{3}-\alpha_{4})}
\end{array}
\right).
\end{equation}

Therefore we have finally
\begin{equation}
\label{eq:last formula}
\mbox{e}^{A}=f_{0}E+f_{1}A+f_{2}A^{2}+f_{3}A^{3}
\end{equation}
with

\begin{eqnarray}
f_{0}&=&
\frac{\alpha_{2}\alpha_{3}\alpha_{4}\mbox{e}^{\alpha_{1}}}
{(\alpha_{2}-\alpha_{1})(\alpha_{3}-\alpha_{1})(\alpha_{4}-\alpha_{1})}
+
\frac{\alpha_{1}\alpha_{3}\alpha_{4}\mbox{e}^{\alpha_{2}}}
{(\alpha_{1}-\alpha_{2})(\alpha_{3}-\alpha_{2})(\alpha_{4}-\alpha_{2})}
\nonumber \\
&&+
\frac{\alpha_{1}\alpha_{2}\alpha_{4}\mbox{e}^{\alpha_{3}}}
{(\alpha_{1}-\alpha_{3})(\alpha_{2}-\alpha_{3})(\alpha_{4}-\alpha_{3})}
+
\frac{\alpha_{1}\alpha_{2}\alpha_{3}\mbox{e}^{\alpha_{4}}}
{(\alpha_{1}-\alpha_{4})(\alpha_{2}-\alpha_{4})(\alpha_{3}-\alpha_{4})}, \\
f_{1}&=&
-\frac{(\alpha_{2}\alpha_{3}+\alpha_{2}\alpha_{4}+\alpha_{3}\alpha_{4})
\mbox{e}^{\alpha_{1}}}
{(\alpha_{2}-\alpha_{1})(\alpha_{3}-\alpha_{1})(\alpha_{4}-\alpha_{1})}
-
\frac{(\alpha_{1}\alpha_{3}+\alpha_{1}\alpha_{4}+\alpha_{3}\alpha_{4})
\mbox{e}^{\alpha_{2}}}
{(\alpha_{1}-\alpha_{2})(\alpha_{3}-\alpha_{2})(\alpha_{4}-\alpha_{2})}
\nonumber \\
&&-
\frac{(\alpha_{1}\alpha_{2}+\alpha_{1}\alpha_{4}+\alpha_{2}\alpha_{4})
\mbox{e}^{\alpha_{3}}}
{(\alpha_{1}-\alpha_{3})(\alpha_{2}-\alpha_{3})(\alpha_{4}-\alpha_{3})}
-
\frac{(\alpha_{1}\alpha_{2}+\alpha_{1}\alpha_{3}+\alpha_{2}\alpha_{3})
\mbox{e}^{\alpha_{4}}}
{(\alpha_{1}-\alpha_{4})(\alpha_{2}-\alpha_{4})(\alpha_{3}-\alpha_{4})}, \\
f_{2}&=&
\frac{(\alpha_{2}+\alpha_{3}+\alpha_{4})\mbox{e}^{\alpha_{1}}}
{(\alpha_{2}-\alpha_{1})(\alpha_{3}-\alpha_{1})(\alpha_{4}-\alpha_{1})}
+
\frac{(\alpha_{1}+\alpha_{3}+\alpha_{4})\mbox{e}^{\alpha_{2}}}
{(\alpha_{1}-\alpha_{2})(\alpha_{3}-\alpha_{2})(\alpha_{4}-\alpha_{2})}
\nonumber \\
&&+
\frac{(\alpha_{1}+\alpha_{2}+\alpha_{4})\mbox{e}^{\alpha_{3}}}
{(\alpha_{1}-\alpha_{3})(\alpha_{2}-\alpha_{3})(\alpha_{4}-\alpha_{3})}
+
\frac{(\alpha_{1}+\alpha_{2}+\alpha_{3})\mbox{e}^{\alpha_{3}}}
{(\alpha_{1}-\alpha_{4})(\alpha_{2}-\alpha_{4})(\alpha_{3}-\alpha_{4})},\\
f_{3}&=&
-\frac{\mbox{e}^{\alpha_{1}}}
{(\alpha_{2}-\alpha_{1})(\alpha_{3}-\alpha_{1})(\alpha_{4}-\alpha_{1})}
-
\frac{\mbox{e}^{\alpha_{2}}}
{(\alpha_{1}-\alpha_{2})(\alpha_{3}-\alpha_{2})(\alpha_{4}-\alpha_{2})}
\nonumber \\
&&-
\frac{\mbox{e}^{\alpha_{3}}}
{(\alpha_{1}-\alpha_{3})(\alpha_{2}-\alpha_{3})(\alpha_{4}-\alpha_{3})}
-
\frac{\mbox{e}^{\alpha_{3}}}
{(\alpha_{1}-\alpha_{4})(\alpha_{2}-\alpha_{4})(\alpha_{3}-\alpha_{4})}.
\end{eqnarray}

\par \vspace{5mm}
{\bf A comment is in order}.

\par \noindent
(1)\ At first sight, the eigenvalues in our 
formula (\ref{eq:last formula}) appear to be different all. 
However, it is not true. Let us explain this with a simple example. 

For $A \in M(2;\fukuso)$ 
\[
0=|\lambda E-A|=\lambda^{2}+p_{1}\lambda+p_{2}
=(\lambda-\alpha)(\lambda-\beta),
\]
which gives\quad $p_{1}=-(\alpha+\beta),\ p_{2}=\alpha\beta$\quad and 
\[
L=
\left(
\begin{array}{cc}
 0 & -p_{2} \\
 1 & -p_{1}
\end{array}
\right).
\]
Then the formula is 
\[
\mbox{e}^{A}=(E, A)\mbox{e}^{L}
\left(
\begin{array}{c}
 1 \\
 0
\end{array}
\right).
\]
In the following let us consider two cases. 

\par \vspace{3mm} \noindent
(I)\quad $\alpha\ne \beta$ :

\par 
In this case, $L=U_{L}
\left(
\begin{array}{cc}
 \beta &         \\
       & \alpha
\end{array}
\right)
U_{L}^{-1}$ with 
\[
U_{L}=
\left(
\begin{array}{cc}
 -\alpha & -\beta \\
       1 & 1
\end{array}
\right)
\Longrightarrow
U_{L}^{-1}=\frac{1}{\beta-\alpha}
\left(
\begin{array}{cc}
  1  & \beta \\
  -1 & -\alpha
\end{array}
\right)
\]
and it is easy to see 
\begin{equation}
\label{eq:diagonal form}
\mbox{e}^{L}
\left(
\begin{array}{c}
 1 \\
 0
\end{array}
\right)
=
\left(
\begin{array}{c}
 \frac{\beta\mbox{e}^{\alpha}-\alpha\mbox{e}^{\beta}}{\beta-\alpha} \\
 \frac{\mbox{e}^{\beta}-\mbox{e}^{\alpha}}{\beta-\alpha}
\end{array}
\right)
=
\left(
\begin{array}{c}
 \mbox{e}^{\alpha}-
 \alpha\frac{\mbox{e}^{\beta}-\mbox{e}^{\alpha}}{\beta-\alpha} \\
 \frac{\mbox{e}^{\beta}-\mbox{e}^{\alpha}}{\beta-\alpha}
\end{array}
\right).
\end{equation}

\par \vspace{3mm} \noindent
(II)\quad $\alpha=\beta$ :

\par 
In this case, $L$ is 
\[
L=
\left(
\begin{array}{cc}
  0 & -p_{2} \\
  1 & -p_{1}
\end{array}
\right)
=
\left(
\begin{array}{cc}
  0 & -\alpha^{2} \\
  1 & 2\alpha
\end{array}
\right),
\]
we cannot diagonalize $L$, so we must use the method of Jordan canonical 
form. Since this method is well--known (see \cite{IS}) the details are 
omitted. If we set 
\[
U_{L}=
\left(
\begin{array}{cc}
 -\alpha & 1 \\
       1 & 0
\end{array}
\right)
\Longrightarrow
U_{L}^{-1}=
\left(
\begin{array}{cc}
  0 & 1      \\
  1 & \alpha
\end{array}
\right)
\]
then it is easy to see
\[
U_{L}
\left(
\begin{array}{cc}
  \alpha & 1      \\
       0 & \alpha
\end{array}
\right)
U_{L}^{-1}
=
\left(
\begin{array}{cc}
  0 & -\alpha^{2} \\
  1 & 2\alpha
\end{array}
\right)
=
L.
\]
Therefore
\[
\mbox{e}^{L}
=
U_{L}
\mbox{exp}
\left\{
\left(
\begin{array}{cc}
  \alpha & 1      \\
       0 & \alpha
\end{array}
\right)
\right\}
U_{L}^{-1}
=
U_{L}
\left(
\begin{array}{cc}
  \mbox{e}^{\alpha} & \mbox{e}^{\alpha} \\
       0            & \mbox{e}^{\alpha}
\end{array}
\right)
U_{L}^{-1},
\]
we have
\begin{equation}
\label{eq:jordan canonical form}
\mbox{e}^{L}
\left(
\begin{array}{c}
 1 \\
 0
\end{array}
\right)
=
\left(
\begin{array}{c}
 (1-\alpha)\mbox{e}^{\alpha} \\
 \mbox{e}^{\alpha}
\end{array}
\right).
\end{equation}

\par \vspace{3mm} \noindent
Comparing (\ref{eq:jordan canonical form}) with (\ref{eq:diagonal form}) 
it is clear that 
$\lim_{\beta\to\alpha}$(\ref{eq:diagonal form})
=
(\ref{eq:jordan canonical form}). That is, our method covers all special 
cases by taking some appropriate limit. 

\par \noindent
To obtain the formula including multi--plicities of eigenvalues is a good 
exercise. We leave it to readers.

\par \vspace{5mm} \noindent
(2)\ In realistic problems we must calculate $\mbox{e}^{tA}$ in place of 
$\mbox{e}^{A}$. The result is changed simply into
\begin{equation}
\mbox{e}^{tA}=f_{0}E+f_{1}A+f_{2}A^{2}+\cdots +f_{n-1}A^{n-1}
\end{equation}
with
\begin{equation}
f_{l}=(-1)^{n+1}\sum_{k=1}^{n}
\frac{\left(p_{n-l-1}\right)_{k}\mbox{e}^{t\alpha_{k}}}{\prod_{j=1,j\ne k}^{n}
(\alpha_{j}-\alpha_{k})}\qquad 0\leq l\leq n-1.
\end{equation}
We leave the check to readers.

\par \vspace{5mm} \noindent
(3)\ We comment on a generalization of the result. Let $F$ be an 
entire function on $\fukuso$. Then the result is changed simply into
\begin{equation}
F(A)=f_{0}E+f_{1}A+f_{2}A^{2}+\cdots +f_{n-1}A^{n-1}
\end{equation}
with
\begin{equation}
f_{l}=(-1)^{n+1}\sum_{k=1}^{n}
\frac{\left(p_{n-l-1}\right)_{k}F(\alpha_{k})}{\prod_{j=1,j\ne k}^{n}
(\alpha_{j}-\alpha_{k})}\qquad 0\leq l\leq n-1.
\end{equation}
We leave the check to readers.

\par \vspace{10mm}
We conclude this paper by stating our motivation. 
We are studying a quantum computation based on Cavity QED whose image is

\vspace{10mm}
%%%%%%%%%%%%%%%%%%%%%%%%%%%%%%%%%%%%%%%%%%%%%%%%%%%%%%%%%%%%%%%%%%%%
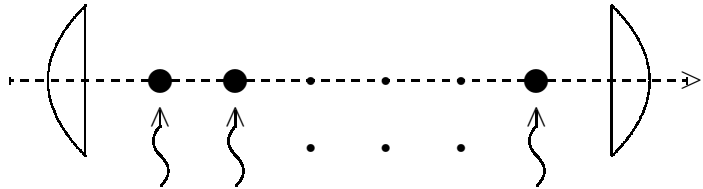
\begin{center}
\setlength{\unitlength}{1mm} 
\begin{picture}(110,40)(0,-20)
\bezier{200}(20,0)(10,10)(20,20)
\put(20,0){\line(0,1){20}}
\put(30,10){\circle*{3}}
\bezier{200}(30,-4)(32,-2)(30,0)
\bezier{200}(30,0)(28,2)(30,4)
\put(30,4){\line(0,1){2}}
\put(28.6,4){$\wedge$}
\put(40,10){\circle*{3}}
\bezier{200}(40,-4)(42,-2)(40,0)
\bezier{200}(40,0)(38,2)(40,4)
\put(40,4){\line(0,1){2}}
\put(38.6,4){$\wedge$}
\put(50,10){\circle*{1}}
\put(60,10){\circle*{1}}
\put(70,10){\circle*{1}}
\put(50,1){\circle*{1}}
\put(60,1){\circle*{1}}
\put(70,1){\circle*{1}}
\put(80,10){\circle*{3}}
\bezier{200}(80,-4)(82,-2)(80,0)
\bezier{200}(80,0)(78,2)(80,4)
\put(80,4){\line(0,1){2}}
\put(78.6,4){$\wedge$}
\bezier{200}(90,0)(100,10)(90,20)
\put(90,0){\line(0,1){20}}
\put(10,10){\dashbox(90,0)}
\put(99,9){$>$}
\end{picture}
\end{center}
\vspace{-20mm}
\begin{center}
{The general setting for a quantum computation based on Cavity QED : \\
the dotted line means a single photon inserted in the cavity and \\
all curves mean external laser fields subjected to atoms}
\end{center}
%%%%%%%%%%%%%%%%%%%%%%%%%%%%%%%%%%%%%%%%%%%%%%%%%%%%%%%%%%%%%%%%%
\vspace{5mm}
See \cite{FHKW1}, \cite{FHKW2} in detail. 
It is usually based on two level system of atoms. However, to take a multi 
level system of them into consideration may be better from the view point of 
decoherence which is a severe problem in quantum computation. 
To develop quantum circuits (see for example \cite{KF4}, \cite{FFK} and 
\cite{RZ}) we often encounter the problem to calculate the exponential of 
matrices in an explicit manner, which was very difficult (for at least Fujii). 

Since we have somewhat overcomed this difficulty it must be possible to 
reconsider quantum circuits in the multi level system.

\vspace{10mm}
\noindent{\em Acknowledgment.}\\
K. Fujii wishes to thank Akira Asada, Kunio Funahashi, Shin'ichi Nojiri and 
T. Suzuki for their helpful comments and suggestions.

%%%%%%%%%%%%%
%References%
%%%%%%%%%%%%%

\end{document}